\documentclass[twocolumn]{aastex61}

\usepackage{graphicx}
\usepackage{amssymb}
\usepackage{tabularx}
\usepackage{mathrsfs}
\usepackage{float}
\usepackage{array}
\usepackage{multirow}
\usepackage[colorinlistoftodos]{todonotes}
\usepackage{xspace}
\usepackage{hyperref}

\newcommand{\Nmin}{\ensuremath{N_{\rm min}}}
\newcommand{\Mmin}{\ensuremath{M_{\rm min}}}

\newcommand{\hmsun}{\ensuremath{h^{-1}\;M_\odot}}
\newcommand{\hmpc}{\ensuremath{h^{-1}\;{\rm Mpc}}}
\newcommand{\ihmpc}{\ensuremath{h\;{\rm Mpc}^{-1}}}
\newcommand{\abacus}{\textsc{Abacus}\xspace}

\begin{document}

\title{Testing the Detection Significance on the Large Scale Structure by a JWST Deep Field Survey}

\author{Hao Zhang}
\affiliation{School of Physics, Peking University, Beijing 100871, China}
\affiliation{Harvard-Smithsonian Center for Astrophysics, Harvard University, Cambridge 02138, MA, USA}
\author{Daniel J.\ Eisenstein}
\author{Lehman H.~Garrison}
\author{Douglas W.~Ferrer}
\affiliation{Harvard-Smithsonian Center for Astrophysics, Harvard University, Cambridge 02138, MA, USA}

\begin{abstract}
In preparation for deep extragalactic imaging with the James Webb Space Telescope, we explore the clustering of massive halos at $z=8$ and $10$ using a large N-body simulation.  We find that halos with masses $10^9$ to $10^{11}$~$\hmsun$, which are those expected to host galaxies detectable with JWST, are highly clustered with bias factors ranging from 5 and 30 depending strongly on mass, as well as on redshift and scale. This results in correlation lengths of 5--10\hmpc, similar to that of today's galaxies. Our results are based on a simulation of 130 billion particles in a box of $250\hmpc$ size using our new high-accuracy \abacus simulation code, the corrections to cosmological initial conditions of \citet{2016MNRAS.461.4125G}, and the \textit{Planck} 2015 cosmology.  We use variations between sub-volumes to estimate the detectability of the clustering.  Because of the very strong inter-halo clustering, we find that surveys of order 25\hmpc\ comoving transverse size may be able to detect the clustering of $z=8$--10 galaxies with only 500-1000 survey objects if the galaxies indeed occupy the most massive dark matter halos.

\end{abstract}


\section{Introduction}

Galaxy formation is strongly influenced by large-scale structure. In dark matter halos of high mass, gas is easier to cool and can thus form stars and galaxies \citep{1997MNRAS.286..795K}.
The halos hosting luminous galaxies at high redshift are expected to be massive, rare, and therefore highly clustered. This in turn implies that the galaxies should be highly clustered, corresponding to large bias values \citep{1999MNRAS.303..188K, 2017MNRAS.469.4428J}. Observations on Lyman $\alpha$ emitters \citep{2014PASJ...66R...1T,  2015MNRAS.453.1843S, 2017arXiv170407455O}, Lyman $\alpha$ blobs \citep{2006A&A...452L..23N, 2009ApJ...693.1579Y,  2010ApJ...719.1654Y}, Lyman break galaxies \citep{2014ApJ...793...17B, 2016ApJ...821..123H, 2017arXiv170406535H} and HST deep field observations \citep{2006ApJ...648L...5O, 2013ApJ...768..196S, 2014ApJ...796L..27R} have supported this, finding large angular clustering
and field-to-field density variations.

As massive halos are extreme fluctuations in the density field, the resulting number of these host sites and their clustering is unusually sensitive to the cosmological model. Therefore, measuring the clustering can be indicative for the halo mass function.  This has been a common application of halo occupation distribution modeling, a method in which the association of galaxies to halos of a given mass leads to detailed predictions of galaxy clustering \citep[for a review, see][]{2002PhR...372....1C}.

However, the extreme sensitivity of the clustering on the cosmological properties also requires careful control of the initial conditions and numerical methods.
In this paper, we present a large-scale high-resolution N-body simulation to investigate halo clustering at high redshifts.  Our work includes several improvements that we argue will improve the reliability of the results. First, we adopt our cosmological parameters from the most recent Planck measurements \citep{2016A&A...594A..13P}.  The increase in the matter density, $\Omega_m h^2$, relative to previous CMB results increases the small-scale fluctuations in LCDM and hence the abundance of halos at a given mass.  Second, we utilize our new N-body cosmological code \abacus, which features high force accuracy.  We adopt a small particle mass of $10^7\hmsun$ so that halos of $10^{10}\hmsun$, which we expect will be typical of detectable galaxies with JWST, will be well-resolved.  The high speed of Abacus allows us to still run a box of 130 billion particles filling $248.8\hmpc$, big enough to capture most of the large-scale modes relevant to the formation of these halos.  Third, we utilize the corrections to the linear-theory initial conditions highlighted by \citet{2016MNRAS.461.4125G} and also use the methods in that paper to include second-order perturbation theory terms in the initial conditions.

In order to compare with upcoming JWST deep field surveys at high redshifts, we analyze time slices of our simulation at $z=10$ and $z=8$.  Since the simulation volume of around $(250\hmpc)^2$ is much larger than the volume of a typical JWST survey, we are able to cut the simulation into many sub-volumes and use the variations between them to estimate the covariance of the clustering.  Here, we choose to divide the region into 10 $\times$ 10 boxes, investigate the clustering in our simulated halo catalogs, and predict a detection significance for one box representing a single survey.

While we were preparing this publication, the work of \cite{2017arXiv170702312B} was published which investigated clustering at $z > 7$ using a cosmological hydrodynamical simulation called the BLUETIDES simulation \citep{2016MNRAS.455.2778F}. Their research obtained results that are compatible with ours.

In Section 2, we introduce the simulation used in this work.  In Section 3, we describe our methodology and then present our results. In section 4, we give our conclusion and a discussion.

\section{Cosmological Simulation}

\abacus is a code for cosmological N-body simulation (Ferrer et al., in prep.; Metchnik \& Pinto, in prep.) that is both extremely fast and highly accurate, aided by recent computational techniques and commodity hardware for high performance computing.   Abacus utilizes a novel fully-disjoint split between the near-field and far-field gravitational sources, solving the former on GPU hardware and the latter with a variant of a multipole method.  The result is very high speed, in excess of 20 million particle updates per second on a single 24-core workstation.  Further, Abacus is built to store most of its data on a high-speed disk system, allowing us to run multi-terabyte problems on a single computer with only modest amounts of RAM.

In this paper, we use a single $5120^3$ simulation of a $(248.8\hmpc)^3$ box.  This results in a particle mass of $10^7\hmsun$, suitable to robustly identify halos with masses around $10^{10}\hmsun$.  We evolve the simulation using a standard leap-frog integration with 225 time steps from $z=199$ to $z=10$ and 67 more to $z=8$.  All particles have the same time step.  The simulation was run on a single commodity-based 24-core dual Xeon workstation with 256 GB of RAM, 2 NVidia GeForce GTX 980 Ti GPUs, and a RAID system providing over 1.5 GB/sec of disk speed, with each time step taking about 2.2 hours.

\cite{PhysRevD.73.103507} showed that solutions to the discrete N-body problem do not correctly recover the continuum linear perturbation theory found in cosmological textbooks for wavenumbers near the Nyquist wavenumber.  Most Fourier modes grow too slowly, although a few grow too quickly.  While the effects are small for modes much larger than the inter-particle spacing, we are nevertheless concerned that the formation of extreme halos is very sensitive to small changes in perturbation amplitude.  

We therefore use the initial conditions proposed by \cite{2016MNRAS.461.4125G}, whose method seeks to cancel out these linear theory errors at a given target redshift (here chosen to be $z=49$).  This method is careful to use only the longitudinal linear-theory growing mode, which differs from the wavevector in the discrete theory.  It then adjusts the initial displacement
amplitudes of each mode so as to compensate for the non-standard growth function that will be encountered between the initial redshift of $z=199$ and the target redshift.  Finally, we include second-order effects on the initial perturbations by inverting the particle displacements and using the sum of the forces in both cases to isolate the second-order forces, which are then applied as displacements and velocities assuming the continuum limit.  Such second-order corrections are known to be important for the formation of the most massive halos \citep{doi:10.1111/j.1365-2966.2006.11040.x,2015PhDT.......115S}.

We adopted the cosmology of $\Omega_m = 0.31415$, $\Omega_{DE} = 1 -  \Omega_{m}$, $\Omega_{K} = 0$, $h = 0.6726$ from the Planck measurements \citep{2014A&A...571A..16P}. The linear power spectra was calculated by the package of ``Code for Anisotropies in the Microwave Background'' (CAMB) \citep{2000ApJ...538..473L}. We started at the initial redshift of $z=199$ and used the output time slice at $z = 10$ and $8$.  

The group finding algorithm that we adopt is the Friends-of-Friends algorithm \citep{1982ApJ...259..449P,1982ApJ...257..423H}, which connects all pairs of particles within a certain critical distance and then identifies clumps of interconnected particles above a certain multiplicity threshold as a halo.
As our goal is to establish a more robust prediction of halo abundance and clustering, we require at least 300 particles in a halo and focus on the case of 1000 particles ($10^{10}\hmsun$) and at redshift of $z=10$ as our fiducial case.  This ensures that halos are robustly found.  For example, \citet{2016MNRAS.461.4125G} found that such multiplicity yielded well-converged results with respect to particle discreteness when using the initial conditions developed in that work.

We make the halo catalogs from our simulation available at \url{http://nbody.rc.fas.harvard.edu/public/JWST_products/}. Further documentation of the data files are given in \citep{2017ApJS...Submit...G} and at \url{https://lgarrison.github.io/AbacusCosmos/}.

\section{Large Scale Structure of High-Redshift Halos}

\subsection{Clustering Methodology}

\subsubsection{Two-point Clustering Statistics}\label{2PCFmethod}

We aim to study the clustering of halos as a function of their mass using the two-point clustering statistics: the familiar two-point correlation function (2PCF) and power spectrum.
We define samples based on thresholds in halo mass and compare the results between different threshold values, as well as to the correlations of the matter field and of linear theory.  It is worth noting that halos of the requisite mass are treated as containing only one galaxy.  Halo occupation distribution models commonly assign additional satellite galaxies to the most massive halos, which can further increase the clustering strength, particularly at intra-halo separations but also at inter-halo separations.

For our analysis of the halo clustering, we split the simulation volume into 100 rectangular pieces, each 25 by 25 by 250$h^{-1}$ comoving Mpc.  We introduce the sub-volumes so that we can use the dispersion among the sub-volumes to determine the covariance matrix of the 2PCF.  However, it is also the case that these volumes correspond to roughly the scale of a substantial JWST survey, about 13$'$ wide and $\Delta z=2$ at these redshifts. 

We compute the 2PCF in each sub-volume, ignoring any periodicity,
using the \citet{1993ApJ...412...64L} estimator 
\begin{equation}
\label{eq:xiDR}
\xi(\vec{r}) = \frac{DD - 2DR + RR}{RR}
\end{equation}
where DD, DR and RR indicate the counts in each separation bin of data-data, data-random and random-random halo pairs, respectively.  The random catalog is a uniform distribution across the entire volume.  In detail, we use the simplicity of the rectangular volume to accelerate the DR and RR calculations by building interpolative functions to return the volumes of spheres near to the boundaries.  We confirm that the mean of the sub-volume results is very similar, save at the largest separations, to the result for the full periodic simulation volume, where the DR and RR counts are trivially computed in the infinite sampling limit.  

For the correlations of the non-linear matter field and linear theory, we first obtain their power spectra and then compute the 2PCFs from the power spectra based on the inverse Fourier Transform relation described in Eq.~\ref{eq:xi}.
\begin{equation}
\label{eq:xi}
\xi(r) = \int dk \frac{k^2}{2\pi^2}  \frac{\sin(kr)}{kr} e^{-(k\Sigma)^2}P(k) 
\end{equation}

The power spectra of the halos and of the matter field has been calculated in the conventional way using Fourier transforms
of a large periodic gridded representation of the density field.  Shot noise is removed as presented in \citet{2015MNRAS.453L..11B} and we divide by the transfer function of the grid with aliasing \citep{2005ApJ...620..559J}.

\subsubsection{Detection Significance}
Based on the 2PCFs of the 100 sub-volumes, the $(j, k)$ entry of the covariance matrix $\mathbf{C}_{jk}$ is given by
\begin{equation}
\mathbf{C}_{jk} = \frac{1}{N - 1}\sum_{i = 1}^{N} \mathbf{d}_{ij} \mathbf{d}_{ik},
\label{eq:cov}
\end{equation}
where $\mathbf{d}_{ij}$ denotes the j-th separation bin of the 2PCF in the i-th sub-volume.
We then compute the detection significance by
\begin{equation}
\chi^2 = \sum_{i = 1}^{n}\sum_{j = 1}^{n}(\mathbf{d}_{obs, i} - \mathbf{d}_{mean, i})(\mathbf{{C}}^{-1})_{ij}(\mathbf{d}_{obs, j} - \mathbf{d}_{mean, j})
\label{eq:chi2}
\end{equation}
where $\mathbf{d}_{mean, i} = \frac{1}{N}(\sum_{k = 1}^{N} \mathbf{d}_{k,i})$. We use $\mathbf{d}_{obs,i} = 0$ to correspond to the unclustered case, which we interpret as a non-detection.

\subsection{Results}

\subsubsection{Halo Sample Overview}

\begin{table*}[t]
\caption{We investigate the effects of changing the particle number cut value for $z=10$ halos. For each \Nmin, we examine the number of halos in our sample, the 3D and 2D 2PCFs at two representative distances,  the power spectra at two representative lengths of wave vectors and the  3D and 2D $\chi^2$ detection significance for a 1\% subvolume of our simulation.  }
\centering
\label{propsbyN}
\begin{tabular}{|ccc|cccccc|cc|}
\hline
\small{$\Nmin$} & \small{$\Mmin$}  & \small{Number} 
& \multicolumn{2}{c}{$\xi(r)$}
& \multicolumn{2}{c}{\small{$w(R)$}}
& \multicolumn{2}{c|}{\small{$P(k)$ ($h^{-3}\;{\rm Mpc}^3$)}}
& \small{$\chi_{3D}^2$} & \small{$\chi_{2D}^2$}\\ 
& ($10^9\hmsun$) & of halos & $1\hmpc$ & $5\hmpc$ & $1\hmpc$ & $5\hmpc$ & 0.1\ihmpc & 1\ihmpc & & \\
\hline 
300 & 3.0 &  296364 & 10.9 & 0.85 & 0.19 & 0.060 & 1996  &  185 & 85 & 52 \\ 
450 & 4.5 &  138720 & 15.2 & 1.04 & 0.25 & 0.070 & 2468  &  257 & 51 & 27 \\ 
700 & 7.0 &  57127 & 22.5 & 1.31 & 0.35 & 0.087 &  3190  & 380 &   27 & 20 \\ \hline
1000 & 10.0 &  26864 & 32.4 & 1.61 & 0.45 & 0.106 &  3976 &  536 & 19 & 15 \\ \hline
1500 & 15.0 &  10668 & 52.4 & 2.02 & 0.64 & 0.130 &  5250 &  863 & 10 & 8 \\ 
2000 & 20.0 &  5341 & 76.6 & 2.34 & 0.88 & 0.149 &  6283 &  1246 & 5 & 4 \\ \hline
\multicolumn{3}{|c|}{Matter Density Field} & $8.25 \times 10^{-2}$ & $1.43 \times 10^{-2}$  & \nodata & \nodata & 83.3 & 1.017 & \nodata & \nodata \\
\multicolumn{3}{|c|}{Linear Theory} & $7.04 \times 10^{-2}$ & $1.36 \times 10^{-2}$ & \nodata & \nodata & 77.8 & 0.931 & \nodata & \nodata \\
\hline
\end{tabular}
\end{table*}

\begin{table*}[t]
\caption{The same as Table \ref{propsbyN}, but at redshift $z=8$. The maximum of the low mass cutoff is extended to $8 \times 10^{10}$ $\hmsun$ so that the sample size of the most massive halos sample remains around 5,000.}
\centering
\label{propsbyN8}
\begin{tabular}{|ccc|cccccc|cc|}
\hline
\small{$\Nmin$} & \small{$\Mmin$}  & \small{Number} 
& \multicolumn{2}{c}{$\xi(r)$}
& \multicolumn{2}{c}{\small{$w(R)$}}
& \multicolumn{2}{c|}{\small{$P(k)$ ($h^{-3}\;{\rm Mpc}^3$)}}
& \small{$\chi_{3D}^2$} & \small{$\chi_{2D}^2$}\\ 
& ($10^9\hmsun$) & of halos & $1\hmpc$ & $5\hmpc$ & $1\hmpc$ & $5\hmpc$ & 0.1\ihmpc & 1\ihmpc & & \\
\hline 
300 & 3.0 &  1916736 & 4.8 & 0.53 & 0.11 & 0.043 & 1201  &  84 & 199 & 105 \\ 
450 & 4.5 &  1038761 & 6.0 & 0.62 & 0.13 & 0.047 & 1412  &  106 & 160 & 117 \\ 
700 & 7.0 &  515064 & 8.0 & 0.74 & 0.16 & 0.055 &  1703  & 139 &   119 & 62 \\ \hline
1000 & 10.0 &  284623 & 10.3 & 0.86 & 0.19 & 0.061 &  2003 &  178 & 99 & 50 \\ \hline
1500 & 15.0 &  140220 & 14.3 & 1.07 & 0.24 & 0.069 &  2488 &  243 & 63 & 38 \\ 
2000 & 20.0 &  82914 & 18.3 & 1.23 & 0.30 & 0.077 &  2913 &  309 & 48 & 26 \\
3000 & 30.0 &  38051 & 27.3 & 1.55 & 0.42 & 0.097 &  3720 &  455 & 27 & 17 \\
4000 & 40.0 &  21267 & 37.6 & 1.83 & 0.52 & 0.114 &  4535 &  628 & 15 & 10 \\
6000 & 60.0 &  8894 & 59.9 & 2.46 & 0.76 & 0.162 &  6184 &  972 & 9 & 6 \\
8000 & 80.0 &  4678 & 84.3 & 3.02 & 1.02 & 0.189 &  8028 &  1392 & 6 & 3 \\\hline
\multicolumn{3}{|c|}{Matter Density Field} & $0.127$ & $2.10 \times 10^{-2}$  & \nodata & \nodata & 113 & 1.58 & \nodata & \nodata \\
\multicolumn{3}{|c|}{Linear Theory} & $0.105$ & $2.03 \times 10^{-2}$ & \nodata & \nodata & 116 & 1.39 & \nodata & \nodata \\
\hline
\end{tabular}
\end{table*}

\begin{figure}[H]
\small
\centering
\includegraphics[width=7cm]{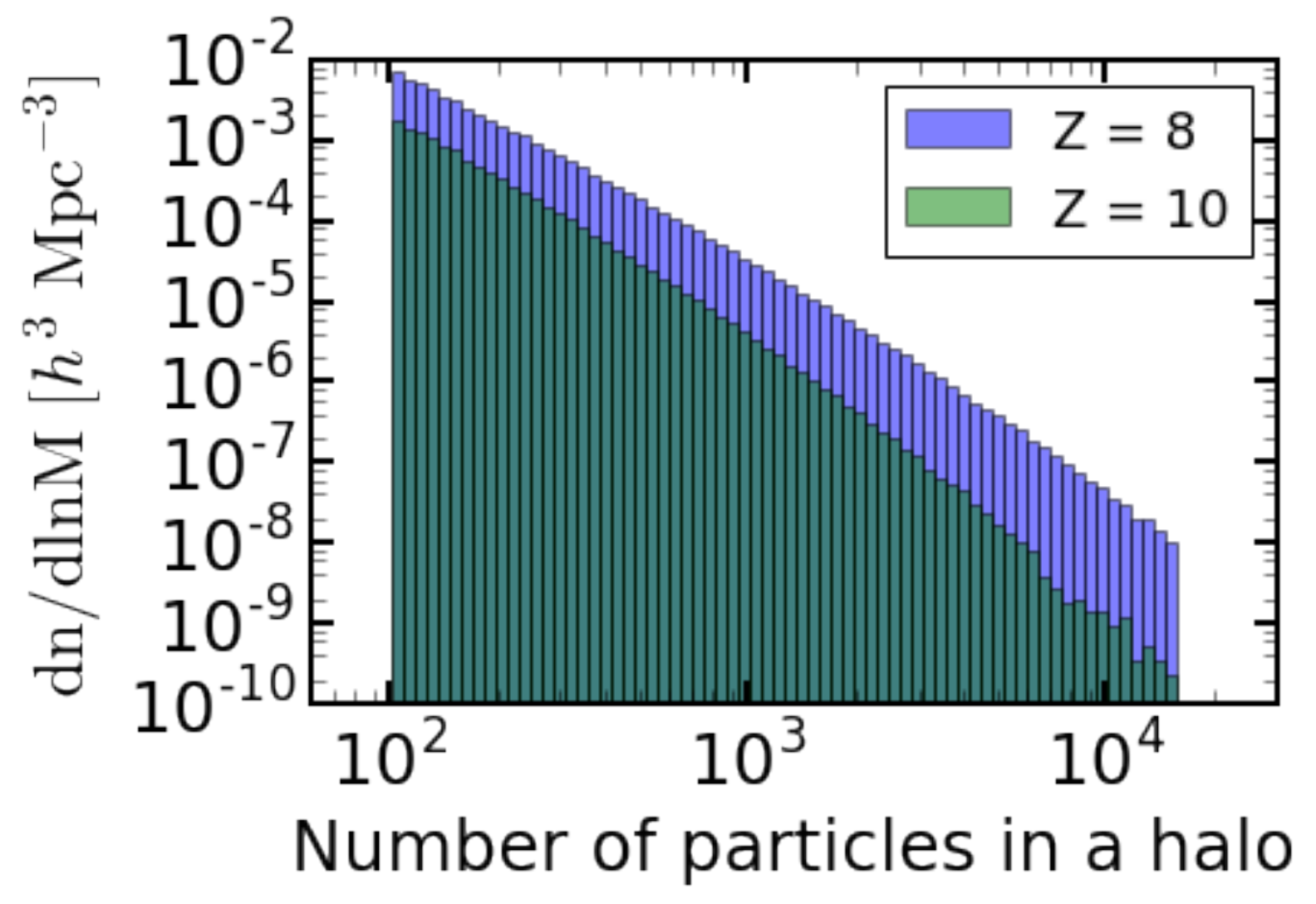}
\caption{A histogram of comoving number density of halos in bins of halo particle multiplicity.  We divide the halo counts by the logarithmic bin width to yield the comoving number density per logarithmic mass bin.  Recall that each particle is $10^7$\hmsun. The two histograms at $z=10$ and $z=8$ are overplotted.}
\label{fig1}
\end{figure}

We begin in Fig. \ref{fig1} with the mass distribution of our halo samples at $z=10$ and $z=8$, obtained by Friends of Friends algorithm.  The number of halos above a series of mass cut are available in Tables \ref{propsbyN} and \ref{propsbyN8}. Below we  primarily use the case of low particle number cut $\Nmin = 1000$ as an illustration.

\begin{figure}[h]
\small
\centering
\includegraphics[width=\linewidth]{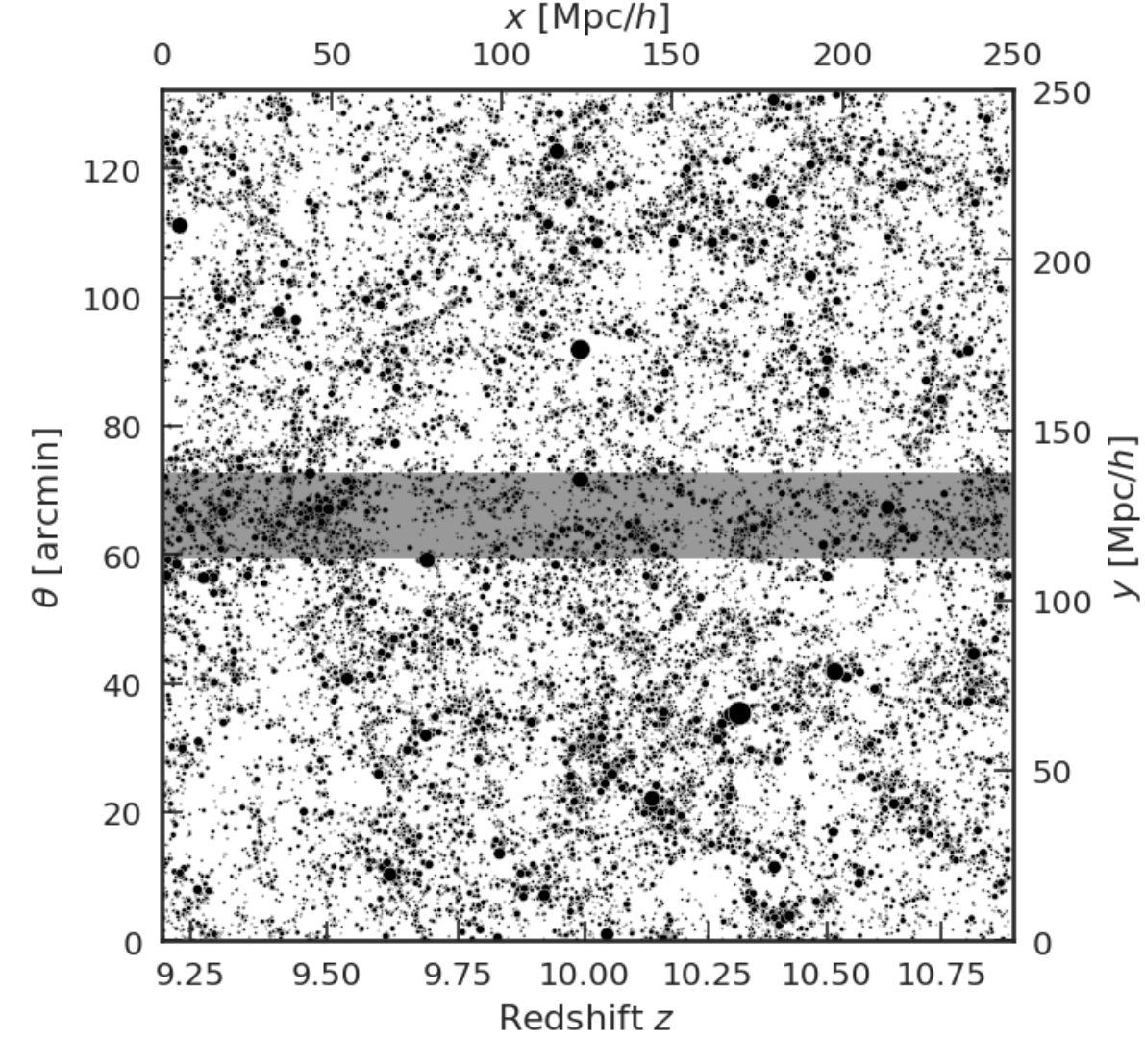}
\caption{A thin slice through the simulation box showing halos larger than 300 particles ($3\times 10^9 \hmsun$) at $z=10$.  Each halo is plotted as a circle with radius proportional to the 90th percentile of the radial particle distribution (``\texttt{r90}''); the radii are inflated by a factor of 10 for plotting purposes.  We imagine this slice as a side-on view of what an observer to the left of the box would see; thus, the horizontal axis is redshift and the vertical axis is angular position.  The depth of the slice is 25 comoving $\hmpc$, or 13.2', which is the size of one of our ``sub-volumes''.  The horizontal shaded region demarcates the same width.}
\label{fig2}
\end{figure}

Fig.~\ref{fig2} shows a 25$\hmpc$ thick slice of our simulation at $z=10$, the thickness chosen to match the width of one of our sub-volumes.  The shaded region shows the same width, which allows one to gauge survey-to-survey variations by eye.  One can see that there will indeed be such variations, depending on the chance intersection of the survey pencil beam with clusters and voids.

\subsubsection{Clustering in 3D Real Space: Halos, Matter Field and Linear Theory}

\begin{figure}
\small
\centering
\includegraphics[width=8cm]{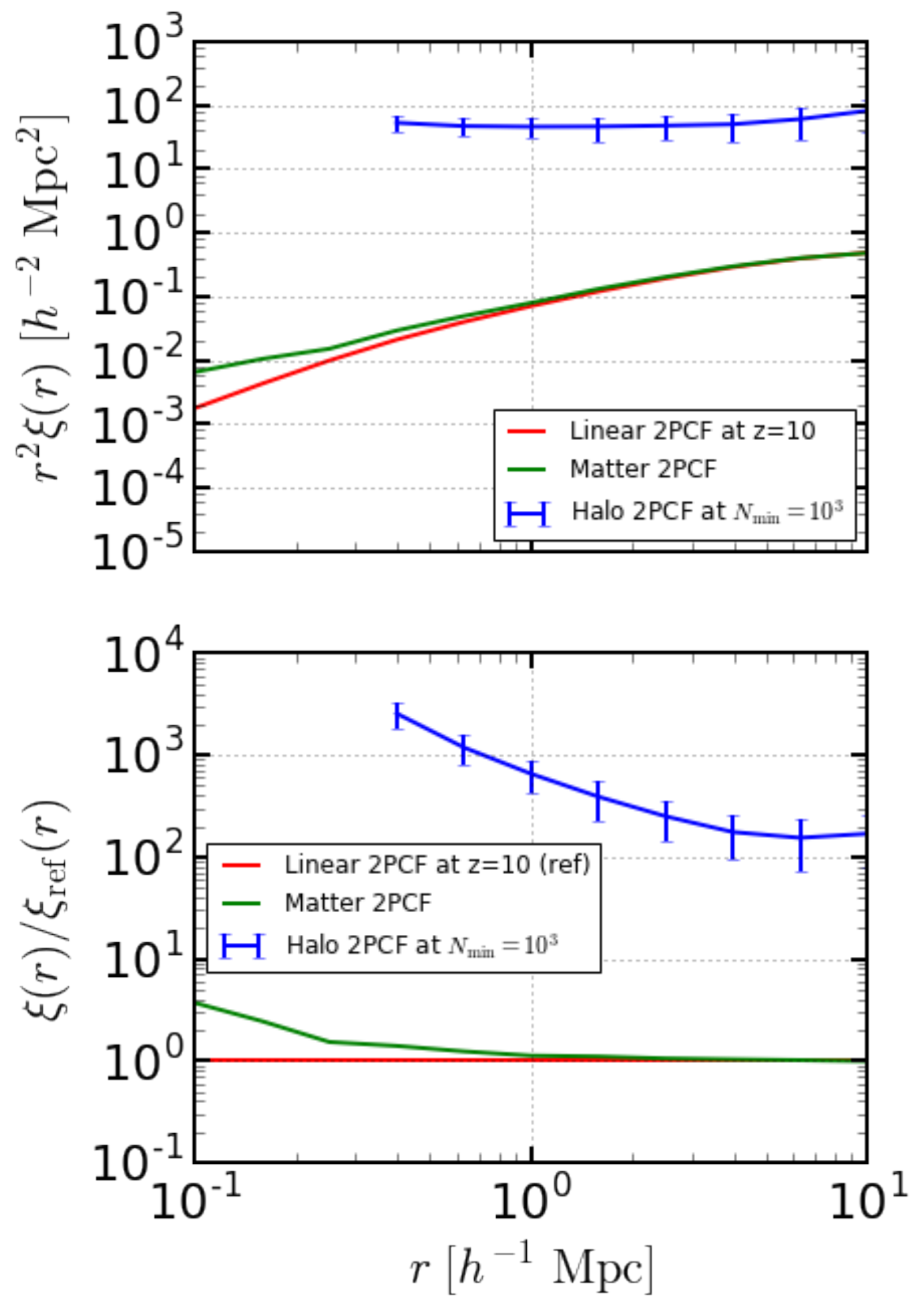}
\caption{Comparison of the matter 2PCF (green), the halo 2PCF (blue), and the 2PCF predicted by linear theory at $z=10$ (red) for halos containing more than 1000 particles in a 1\% sub-field, in $r^2\xi(r)$ (upper panel) and $\xi(r)/\xi_\mathrm{ref}(r)$ (lower panel). Note that the error bars in both cases indicate the standard deviation of a 1\% sub-volume in our $10 \times 10$ partitioning (not the error on the mean for the full simulation volume). The y axis is in $r^2\xi(r)$ in the upper panel, where the flat profile of the halo 2PCF indicates the $r^{-2}$ power law relationship. In the lower panel, we can see that the matter power spectra is basically consistent with the prediction of linear theory, except for the distances below the grid scale where the matter 2PCF gets larger by a factor of 4. The halo 2PCF is highly biased by a multiple of $2 \times 10^2$ to $2 \times 10^3$. }
\label{fig3}
\end{figure}

Following the methods presented in Section \ref{2PCFmethod}, we compute the 2PCF of the $z=10$ halos containing more than $N = 1000$ particles and show the result in Fig.~\ref{fig3} along with the 2PCF of the $z=10$ matter field and linear theory.   We adopt the $\Nmin = 1000$, $\Mmin = 10^{10}$ $\hmsun$ case as our representative one.  This corresponds to 270 objects in a $13' \times 13'$ region at $z=10$. 
To remove the steep scale dependence of the 2PCF, we choose to plot the expression $r^2\xi(r)$ in the upper panel of Fig.~\ref{fig3}.  This choice is common in the low-redshift literature.
As we can see, the matter field 2PCF is consistent with the prediction of the linear theory, while the halo 2PCF is larger by a factor of order $10^2$ to $10^3$, corresponding to clustering biases of 10--30.  

To highlight the scale-dependent bias, we repeat these results in the lower panel of Fig.~\ref{fig3} after dividing by the linear theory correlations function.
This shows a notable increase in bias at scales below 2$\hmpc$ compared to a possible plateau at large scale.  We stress that the comoving diameter of a $10^{10}\hmsun$ halo is about $50h^{-1}$~kpc, so this scale dependence in the bias is occuring well outside the halo scale and indeed beyond even the $300h^{-1}$~kpc scale of the initial Lagrangian volume corresponding to this mass. 
We further note that this scale dependence occurs even though we have omitted satellite galaxies from our analysis.  Indeed, because two halos cannot be closer together than the sum of their radii, our 2PCF drop precipitously at small scales.  
We do not plot results interior to 0.3\hmpc\  so as to comfortably avoid this effect.

\begin{figure}
\small
\centering
\includegraphics[width=8cm]{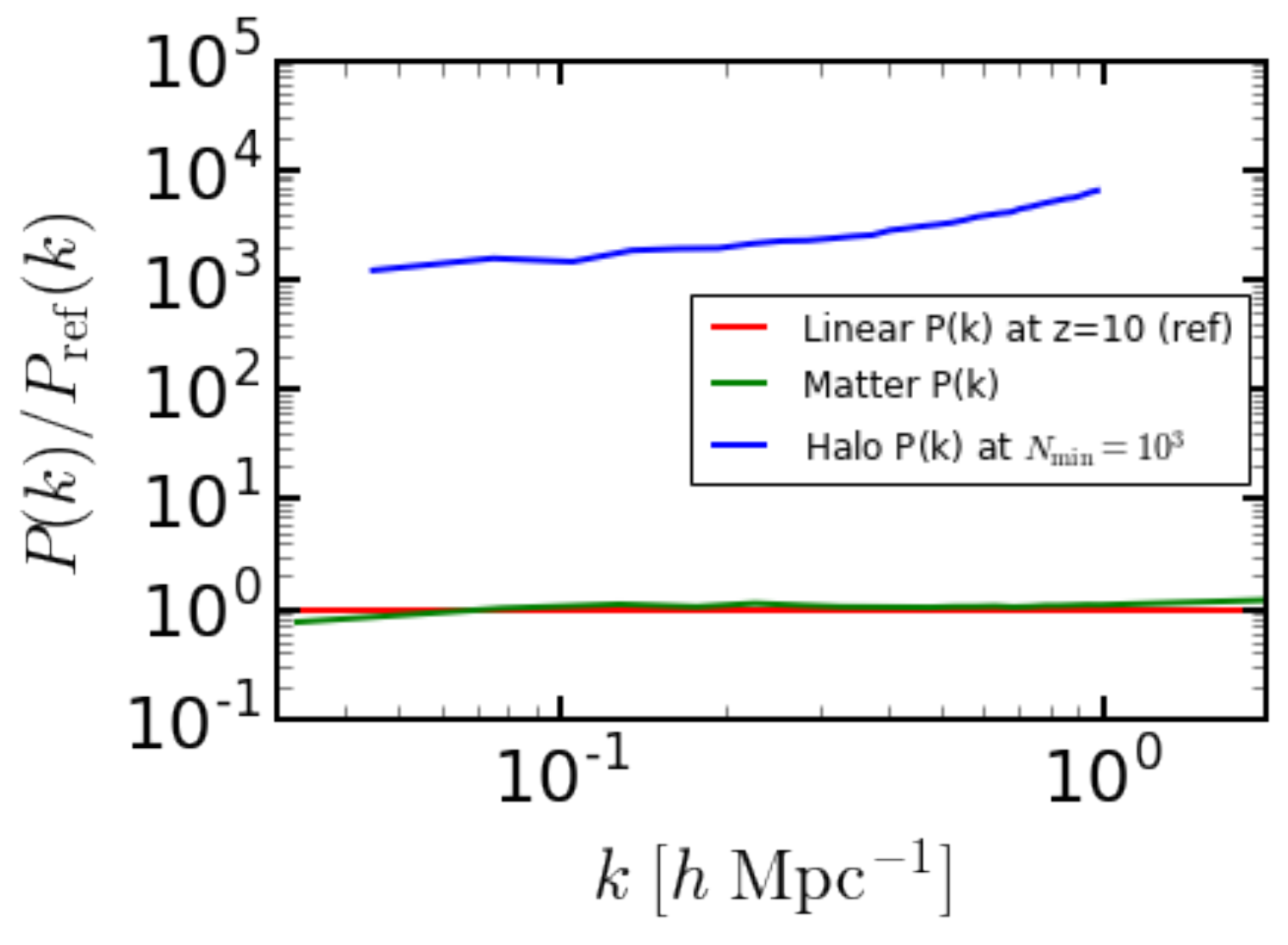}
\caption{Comparison of the matter power spectrum (green), the halo power spectrum (blue), and the power spectrum predicted by linear theory at $z=10$ (red, taken as reference). The matter power spectra is very consistent with the prediction of linear theory. The halo power spectra, however, has a very high bias at the order of $10^2$ to $10^3$.}
\label{fig4}

\end{figure}

We then computed the corresponding power spectra for the three cases. These are shown in Fig. \ref{fig4}, with the linear theory power spectrum being the reference. Again, we obtain a qualitatively similar result as in the previous plot (Fig. \ref{fig3}).

\begin{figure}
\small
\centering
\includegraphics[width=8cm]{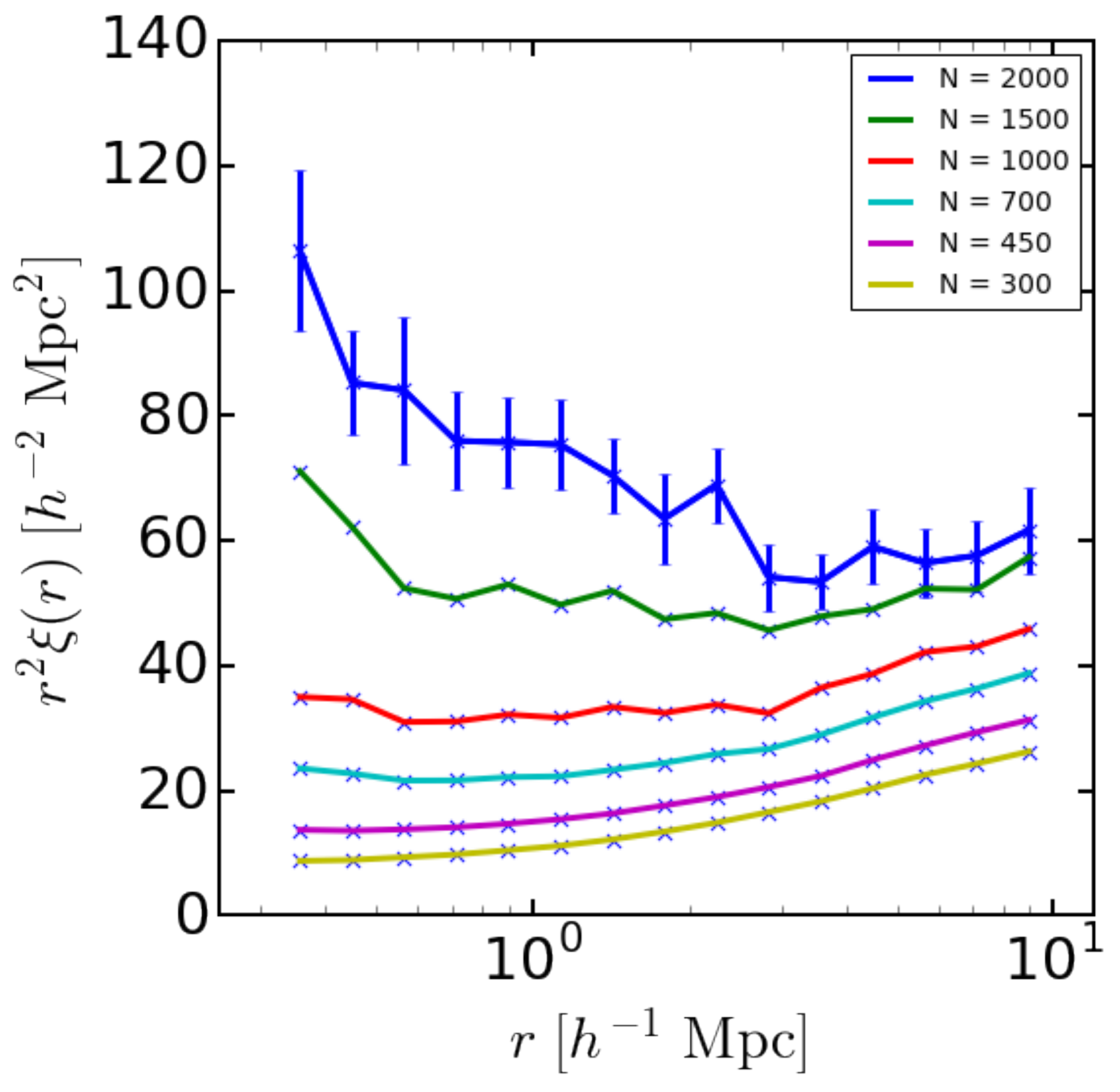}
\caption{The 2PCFs for $z=10$ halos by different mass cutoff represented by the minimum number of particles in the halo finder. For example, the curve labeled 300 shows the 2PCF for all halos that contains more than 300 particles.  One can see the strong trend that the higher mass samples have larger 2PCF amplitudes and higher bias, i.e., that more massive halos are more clustered than less massive halos. For the 2PCF of the highest mass sample, the shot noise in this low number density sample is substantial.  For this sample, we include the standard deviation of the mean 2PCF of our entire simulation box (not as in Fig. \ref{fig3} where the errors for a 1\% sub-volume  were plotted).  The errors for the higher density samples are substantially smaller.  We note that comparisons between curves are partially correlated due to both large-scale structure and the overlapping mass ranges of the halo selections.  This implies that ratios between samples are more tightly constrained than the variance within a sample would suggest.}
\label{fig5}
\end{figure}

\begin{figure}
\small
\centering
\includegraphics[width=8cm]{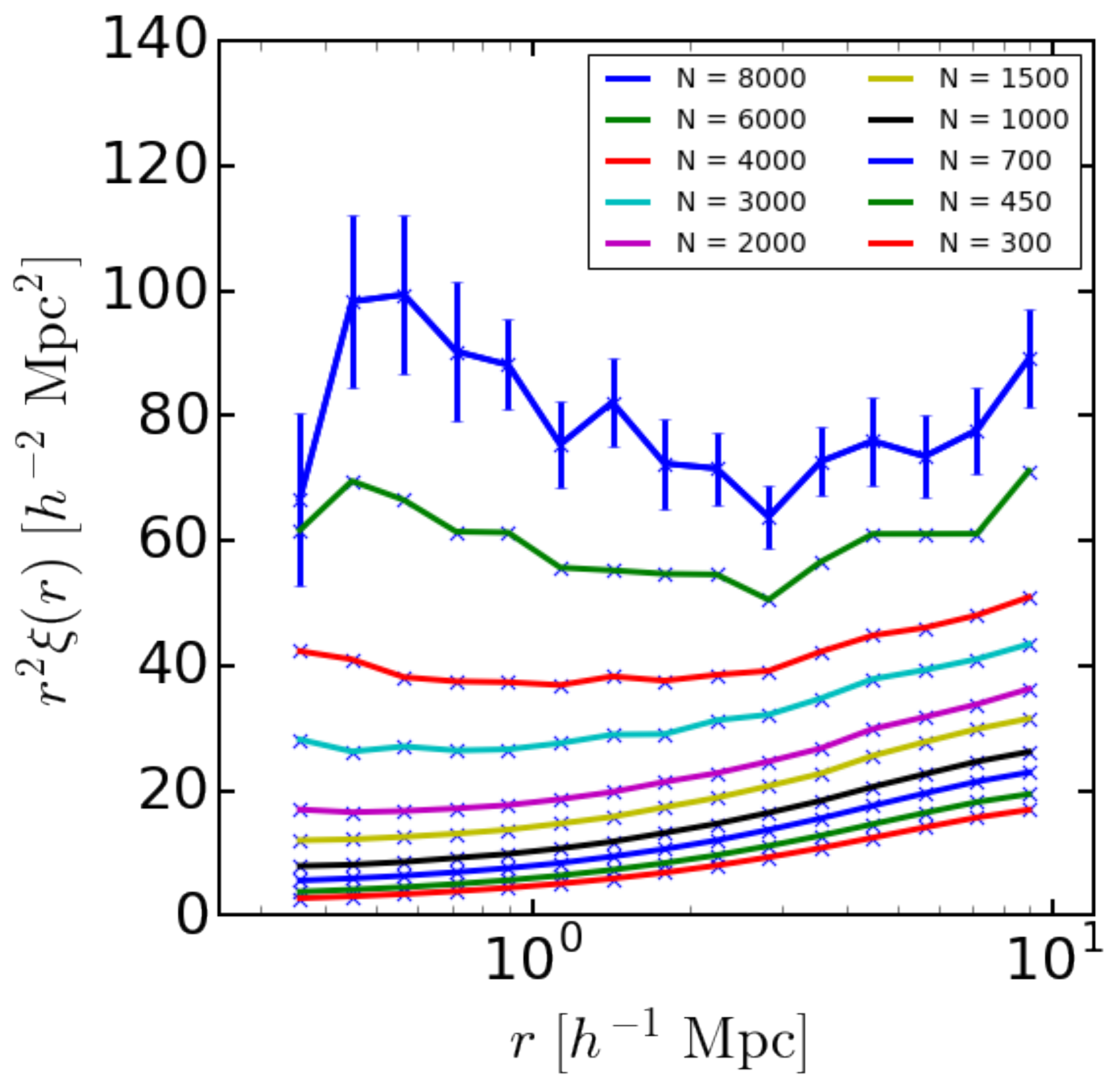}
\caption{The same as Fig. \ref{fig5}, but at redshift $z=8$. We extend the upper limit of the low mass cutoff up to $8 \times 10^{10}$ $\hmsun$ so that the sample size of the most massive halos remains around 5,000 (see Table \ref{propsbyN} and \ref{propsbyN8}).}
\label{fig6}
\end{figure}

Finally we investigate the dependence of the 2PCF on the halo mass cut.  In Fig. \ref{fig5} and \ref{fig6}, we plot $r^2\xi(r)$ at $z=10$ and $z=8$, respectively, for a range of halo mass cuts (3, 4.5, 7, 10, 15, 20 $\times 10^9\ \hmsun$ for both cases, and additionally 30, 40, 60, 80 $\times 10^9\ \hmsun$ for $z=8$).  In Fig. \ref{fig5}, we find a very strong increase in bias for the increasing mass cut.  Further, the correlation functions are shallower than $r^{-2}$ at lower masses, but steeper than $r^{-2}$ at higher masses.  Again, this increase is occuring even though we have not included any satellite galaxies in our catalogs and it involves scales beyond the virialized diameters of the halos.

Fig. \ref{fig6} shows the same progression at $z=8$.  The clustering amplitudes at fixed mass are smaller at low redshift, indicating that clustering bias is falling faster than the growth function is increasing.  However, the clustering amplitudes at fixed number density are more comparable.  Tables \ref{propsbyN} and \ref{propsbyN8} report some characteristic values of different measurements of the clustering.  For the range of the particle number cutoffs mentioned above, we present the various statistics, each at two representative values (1$\hmpc$ and 5$\hmpc$ for the 2PCF, 0.1$\ihmpc$ and 1$\ihmpc$ for power spectrum). As comparisons, we also give the corresponding 2PCFs and power spectrum values obtained from the matter density field and linear theory instead of halos. Comparing the square root of the ratio of the 2PCF indicates bias factor ranging from 5 to 30.

\subsubsection{Halos Clustering in Projected 2D Sky Plane }

In many imaging surveys, our knowledge of the line-of-sight position of galaxies would be limited to the precision of photometric redshifts.  Our measurement of small-scale clustering will then rely on the angular distribution, with the photometric redshifts used to bound the projection effects.  To approximate this situation in our simulation, we project all of the coordinates onto the sky plane by assigning the uniform value to the $x$ coordinate, which is along the redshift direction.  We label the resulting correlation function as $w(r)$, the 2D 2PCF.  Given the $250h^{-1}$ comoving Mpc depth of our box, this corresponds at these redshifts to a projection of about $\Delta z=2$, which is typical of photometric redshift accuracy in Lyman-break samples.

\begin{figure}
\small
\centering
\includegraphics[width=8cm]{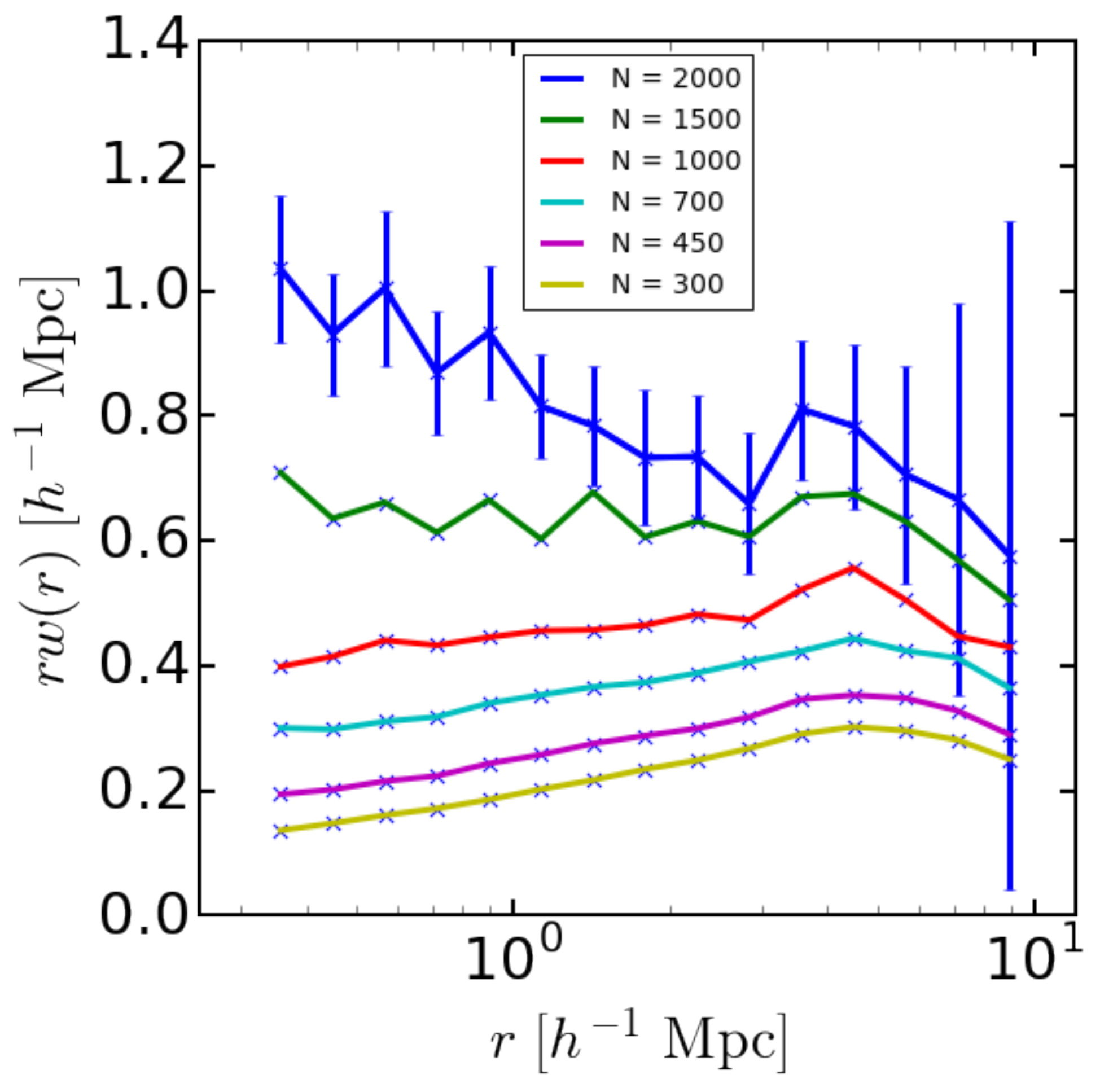}
\caption{The same as Fig. \ref{fig5}, except that this plot shows the 2D 2PCF (projected onto the sky plane) for the corresponding halo particle number cutoffs.}
\label{fig7}
\end{figure}

We investigate the dependence of the 2D 2PCF on the mass cut value. The result is shown in Fig.~\ref{fig7}, which is analogous to Fig.~\ref{fig5} in 3D real space. This time we plot the function of $r w(r)$, equivalent to a $r^{-2}$ correlation function power-law slope.  The stratified structure showing an increasing bias with higher halo mass cut is similar to its 3D counterpart shown in Fig. \ref{fig5}.

\subsubsection{Halos Clustering in 3D Redshift Space}

\begin{figure}
\small
\centering
\includegraphics[width=8cm]{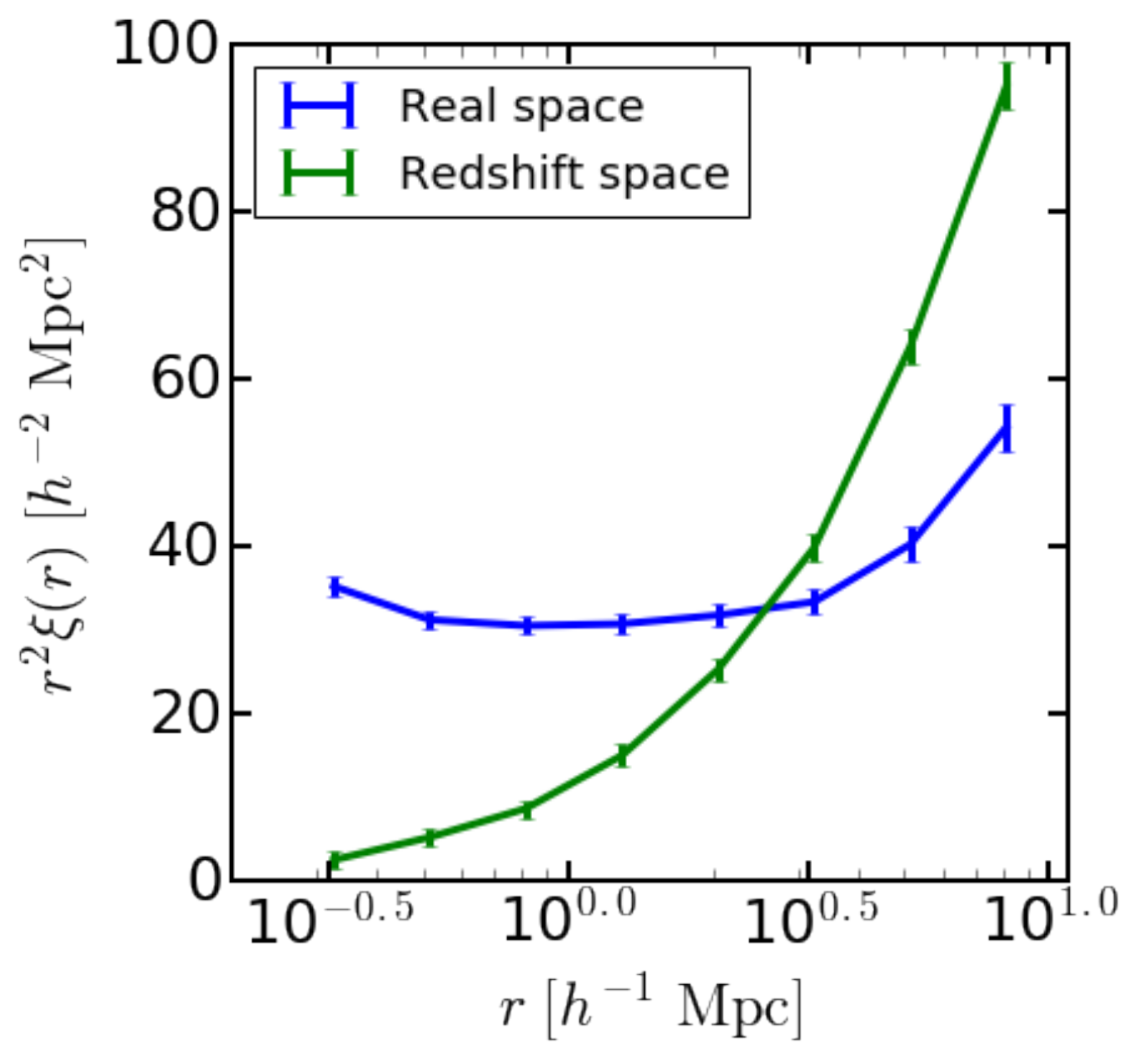}
\caption{The redshift-space 2PCF at $z=10$ for $\Nmin = 1000$ halos compared with the real-space 2PCF of the same sample.  Redshift-space distortions caused by the peculiar velocities of the halo centers of mass bring about a substantial decrease of clustering at small separation and an enhancement at large separations. The error bars are the standard deviations of the mean 2PCFs for the full simulation volume in each case, indicating that the redshift space distortion effect will significantly affect our detection.}
\label{fig8}
\end{figure}

With precise spectroscopic redshifts, one can make more accurate clustering measurements.  In this case, one must contend with the redshift-space distortions caused by peculiar velocities.  Fig. \ref{fig8} shows the 2PCFs in real space and in redshift space for the $\Nmin = 1000$ halos at $z=10$. The distorted 2PCF gives lower correlation on smaller scales and higher correlation on larger scales.  This is expected from the effects of small-scale peculiar velocities, which tends to make nearby objects appear further apart.  

\subsection{Detectability}

We calculate the detectability of these 2PCF based on the covariance matrix derived from the 100 2PCFs corresponding to our 100 sub-volumes. 
The $(i, j)$-th entry of the covariance matrix here is defined as the correlation of $i$-th and $j$-th separation bins in the 2PCF over the 100 sub-volumes; see Eq.~\ref{eq:cov}. The off-diagonal entries are the correlations between two different bins, and the diagonal entries are just variations of each separation bin.  We use 8 bins of radial separation, so as to limit the biases that result from inverting a noisy estimate of the covariance matrix \citep{2014MNRAS.439.2531P}.

\begin{figure}
\small
\centering
\includegraphics[width=8cm]{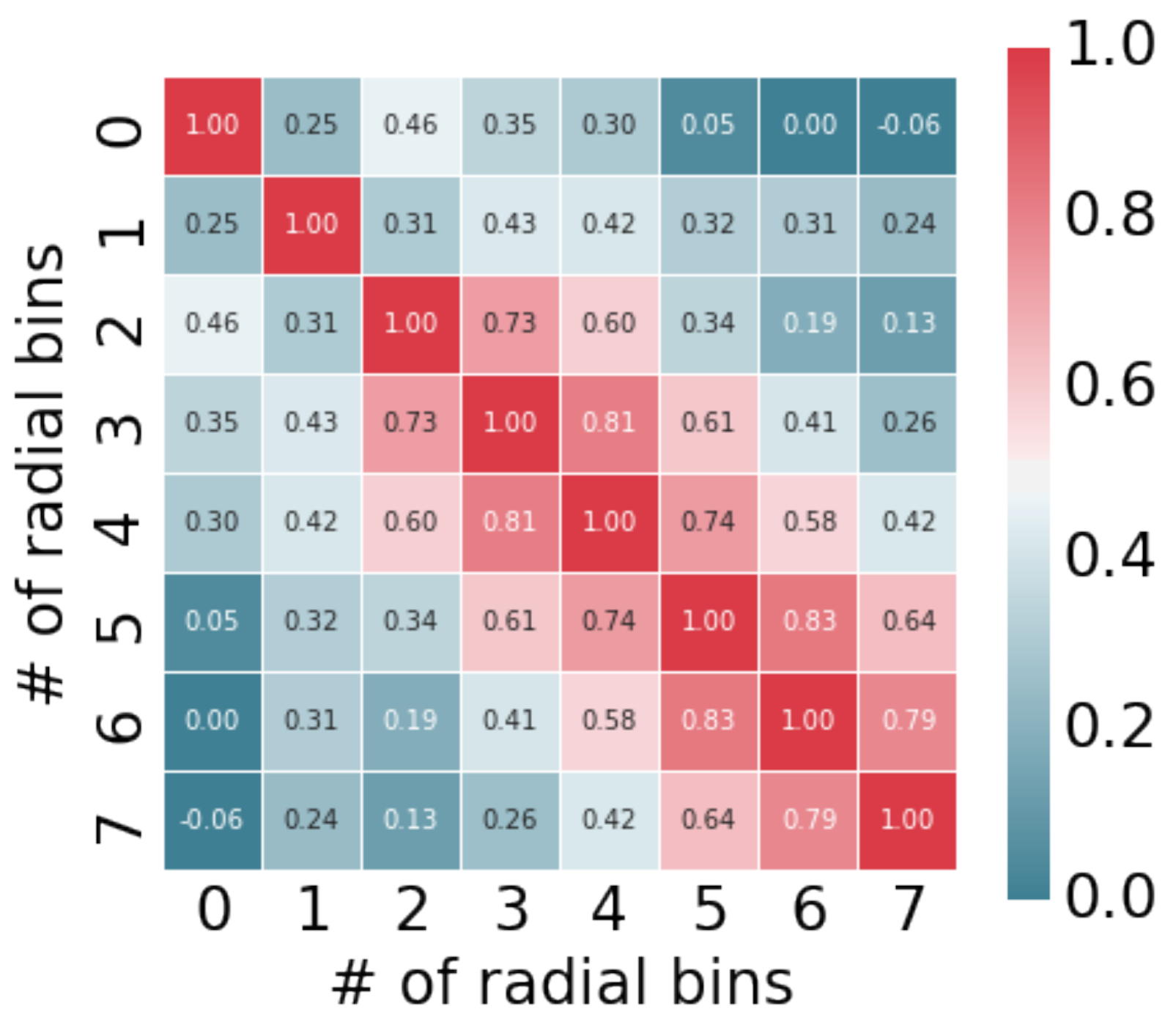}
\caption{The covariance matrix corresponding to the 8 distance bins characterizing the fluctuation of the 2PCFs in 100 sub-volumes in Fig.~\ref{fig3}, where a halo number cutoff of $\Nmin = 1000$ and $z=10$ has been implemented. Each entry is normalized by the formula of $\mathscr{C}_{ij} = C_{ij} / \sqrt{C_{ii}C_{jj}} $, where $\mathscr{C}_{ij}, C_{ij}$ are the normalized and raw entry of the covariance matrix, respectively. Such normalization guarantees all of the diagonal entries to be converted to 1, and all of the off-diagonal entries into the interval of $[-1, 1]$. From this plot we learn that the correlation tends to be larger for closer pair of distance bins, indicating that the 2PCFs for the 100 sub-volumes are fluctuating in a smooth and positively correlated way. Note that the matrix will get closer to a diagonal matrix if we replace $\Nmin = 1000$ with a larger number, indicating the fact that higher $\Nmin$ samples will be more strongly influenced by shot noise which is uncorrelated.}
\label{fig9} 
\end{figure}

In Fig.~\ref{fig9}, we plot an example of the reduced covariance matrix, defined as $\mathscr{C}_{ij} = C_{ij} / \sqrt{C_{ii} C_{jj}}$. From the plot, we see higher correlations for closer separated bins, decaying for progressively farther separated bins. If the variation only consists of Poisson shot noise, then we would expect the variation at each separation bin to be uncorrelated, which is rejected by such a strong off-diagonal covariance. This indicates that there is indeed substantial contribution from the sample variance of large-scale structure  in our 100 sub-volumes.  Repeating this with samples of different mass thresholds shows as expected that the correlations of sparser samples have more diagonally dominated covariances.

We then use the covariance matrix to estimate the detectability of the 2PCF, according to Eq.~\ref{eq:chi2}.  The $\chi^2$ statistic here is equivalently the difference in $\chi^2$ between that of the measured 2PCF and a null $\xi=0$ result.  The interpretation of this in terms of detection significance depends on one's choice of model.  If one had a fully unconstrained model, then one could claim a clustering detection only if the result was unusually large compared to a $\chi^2$ distribution with degrees of freedom equal to the number of bins.  In our case with 8 bins, finding $\chi^2>20$ would be a 99\% confident detection. 
However, it is more cosmologically interesting to investigate smooth models, which sharply limits the number of parameters.  As an extreme, if one's model were simply a rescaling of the observed clustering, then one would have 1 degree of freedom, and the significance would be $\sqrt{\chi^2}$ $\sigma$.  More likely, one would additionally include a power-law slope or other scale-dependent parameter.  The resulting interpretations are hence model dependent, but we suggest that a $\chi^2$ of 
20-25 would be good goal in designing a survey large enough for a first detection of large-scale clustering.

Our results for both angular (2D) and spectroscopic redshift-space (3D) clustering are shown in Tables \ref{propsbyN} and \ref{propsbyN8}.  Note that the $\chi^2$ values refer to a 1\% subvolume, i.e., about 13$'$ square and $\Delta z\approx 2$ deep, while the number of halos refers to the number in the full box.  Studying the results, we find that a first detection of the large-scale correlations could result from an angular survey of 500-1000 galaxies, if the galaxies are indeed populating only the most massive halos.  Adding spectroscopy increases the detection sensitivity by removing noise from projection; however, this is most effective when the samples are denser.  It is important to stress that these results utilize only scales above 300$h^{-1}$ comoving kpc, which is much larger than the virial radius of these halos.  In other words, this is only sensitive to the inter-halo clustering; additional signal from intra-halo (or one-halo) clustering at small separations would boost the detection significance but might be less easily related to the halo mass distribution.

\section{Conclusions}

In this paper, we have investigated the clustering of massive halos at $z=8$ and $10$ using a cosmological N-body simulation. We measured the 2PCFs and power spectra of the halo catalog above a range of cutoff masses and compared them with the same measures for matter field and the prediction of linear theory, finding high values of the clustering bias, typically 10--20. We also measured the angular correlation function by doing a line-of-sight projection and found consistent biases.

We then calculated the detectability of this clustering for an example JWST survey. We set its full volume to $(250 \hmpc)^3$. We divide our full simulation into $10 \times 10$ sub-volumes with equal size and estimate the 2PCF covariance matrix for a single sub-volume.  We then measured the $\chi^2$ of the mean 2PCF relative to the null clustering signal.  Based on the angular correlation function at $z=10$ of a sample exceeding $10^{10}$ $\hmsun$, we derived an expectation of $\chi_{2D}^2 = 15$ relative to a null clustering signal from a sample of 270 galaxies.  With spectroscopic information to remove false pairs from projection, this significance would increase to $\chi_{3D}^2 = 19$ for the 3D redshift-space correlation function.  Hence, we find that the samples of 500--1000 galaxies could yield a detectable large-scale clustering signal ($\chi^2>20$) if indeed the detected galaxies inhabit the most massive dark matter halos.  If the joint distribution of galaxy luminosity (or more precisely, detectability) and halo mass has more scatter, then the typical host halo mass will decrease as will the clustering amplitude.  

These results indicate that the inter-halo clustering of $z\approx8$--10 galaxies could be detectable with achievable sample sizes and that the amplitude of the clustering signal can offer some selection between galaxy formation hypotheses.  However, we remind that our results include only the effect of halo clustering.  Galaxy formation may yet depend on additional effects, such as large-scale radiative feedback and reionization, that could cause additional large-scale clustering.  Distinguishing such signals from those of halo clustering might be possible in the shape of the 2PCF or the signatures of higher-point correlations, but any interpretations of early clustering signals will need to include this caveat.

We next compare the clustering measurement at high redshift presented by \cite{2017arXiv170702312B} obtained from BLUETIDES, a hydrodynamical simulation code that incorporates physics of galaxies, with our clustering measurements from a ABACUS, a pure dark-matter N-body gravitational code. The BLUETIDES analysis gets a bias factor of $10.8 \pm 0.7$ for galaxies, which is consistent with our measurements for dark matter halos. In addition, analyzing the results of these two papers via Halo Occupation Distribution (HOD) modeling helps constraining the galaxy-dark matter halo connection (see Section 3 in \cite{2017arXiv170702312B} for the detailed methods). However, our simulation is purely gravitational on dark matter halos without any assumption on smaller scale physics about galaxies, which thus provides a more robust probe of clustering at high redshifts. Another unique feature of our paper is our focus on detectability of clustering from proposed deep field surveys at high redshift.

Our \abacus code is a robust gravitational N-body cosmology simulation code in the following senses. First, we adopted the latest cosmological parameters from Planck mission (see, for example, \cite{2016A&A...594A..13P}). Second, we adopted an improved set of initial conditions as described in \cite{2016MNRAS.461.4125G}, which only takes longitudinal wave mode, compensates for the non-standard growing factor across the simulated redshift range and take into account second order effects. Therefore, our \abacus code is capable of doing simulations which properly evolve the non-linear fluctuations.

Our investigation clearly reinforces the expectation for upcoming high-redshift surveys that there will be significant field-to-field variations in galaxy populations at  $z \approx 10$. But these variations come with an opportunity, that the clustering signal can be measured with moderately scoped surveys, giving a route to constrain the mass distribution of the host halos of these early galaxies.

We make the halo catalogs from our simulation available at \url{http://nbody.rc.fas.harvard.edu/public/JWST_products/}, so that the simulation can be used for additional analyses of clustering and the generation of JWST mock catalogs.

\section*{Acknowledgement}

We thank Marc Metchnik and Philip Pinto for their contributions to the 
Abacus simulation code.
DJE, LHG, and DWF have been supported by grant AST-1313285 from the National Science Foundation, and DJE is additionally supported as a Simons Foundation Investigator.

\clearpage

\bibliographystyle{apj}
\bibliography{bibliography.bib}

\end{document}